\documentclass[onecolumn,12pt,tightenlines,amsmath,secnumarabic,%
    floatfix,amssymb,aps,nofootinbib,letterpaper, showkeys]{revtex4}
\usepackage{times,graphicx,amsthm}
\usepackage{varioref}
\usepackage{hyperref}

\renewcommand{\subsection}[1]{\medskip\noindent\textbf{#1.} }
\def\marginpar#1{}   

\let\lbl=\label
\def\label#1{\lbl{#1}\ifinner\else\marginpar{\ref{#1} #1}\ignorespaces\fi}

\newcommand{\eps}{\varepsilon}

\renewcommand{\phi}{\varphi}

\newcommand{\tsnnls}{\texttt{tsnnls} }
\newcommand{\TAUCS}{\texttt{TAUCS} }

\graphicspath{{./figs/}}

{\makeatletter
 \gdef\xxxmark{%
   \expandafter\ifx\csname @mpargs\endcsname\relax 
     \expandafter\ifx\csname @captype\endcsname\relax 
       \marginpar{xxx}
     \else
       xxx 
     \fi
   \else
     xxx 
   \fi}
 \gdef\xxx{\@ifnextchar[\xxx@lab\xxx@nolab}
 \long\gdef\xxx@lab[#1]#2{{\bf [\xxxmark #2 ---{\sc #1}]}}
 \long\gdef\xxx@nolab#1{{\bf [\xxxmark #1]}}
}

\newcommand{\LSQR}{\texttt{LSQR} }
\newcommand{\lsqnonneg}{\texttt{lsqnonneg} }
\newcommand{\snnls}{\texttt{snnls} }

\begin{document}

\markboth{Piatek, Cantarella}{\texttt{tsnnls}: A sparse nonnegative least squares solver}
\title{\texttt{tsnnls}: A solver for large sparse least squares problems \\ with non-negative variables}

\author{Jason Cantarella}
\email[Email: ]{cantarel@math.uga.edu}
\affiliation{Department of Mathematics, University of Georgia,
Athens, GA 30602}

\author{Michael Piatek}
\email[Email: ]{piatek@mathcs.duq.edu}
\affiliation{Department of Mathematics and Computer Science, Duquesne University,
Pittsburgh, PA 15282}


\begin{abstract}
The solution of large, sparse constrained least-squares problems is a staple in 
scientific and engineering applications. However, currently available codes
for such problems are proprietary or based on \texttt{MATLAB}. We announce a freely available C implementation of the fast block pivoting algorithm of Portugal, Judice, and Vicente. 
Our version is several times faster than Matstoms' \texttt{MATLAB} implementation of the same algorithm. Further, our code matches the accuracy of \texttt{MATLAB}'s built-in \texttt{lsqnonneg} function. 
\end{abstract}



\maketitle

\section{Introduction}
\label{sec:intro}

The authors were recently faced with the challenge of finding 
a fast solver for the sparse non-negative least-squares problem (NNLS) to embed
in a much larger scientific application. The problem is given by 
\begin{equation}
\label{eqn:nnls}
\min_x \frac{1}{2}|| Ax - b ||^2_2 \quad \mbox{with } x \geq 0
\end{equation}
where $A$ is an $m \times n$ matrix, $x \in \mathbb{R}^n$, $b \in \mathbb{R}^m$ and $m > n$, and
we assume $A$ has full column rank. This is a standard problem in numerical linear algebra (\cite{MR96d:65067,MR97g:65004}) which is handled by a number of commercial libraries (\cite{MR2000h:90104,knitro,TOMLAB}) and by the \texttt{MATLAB}-based Sparse Matrix Toolbox of~\cite{sls}. While these methods work well, their users must incur the overhead of a large math package or the expense and license restrictions of commercial libraries. There does not seem to be a freely available solver for this problem without these disadvantages. This motivated the development of \texttt{tsnnls}, a lightweight ANSI C implementation of the block principal pivoting algorithm of~\cite{MR95a:90059} which matches the accuracy of the \texttt{MATLAB}-based codes and is considerably faster.
The code can be obtained at \url{http://www.cs.duq.edu/~piatek/tsnnls/} or
 \url{http://ada.math.uga.edu/research/software/tsnnls/}.  Users may redistribute the library under the terms of the GNU GPL.

\section{Algorithms}
\label{sec:algorithms}

The following is a summary of our main algorithm as described in~\cite{MR95a:90059}. The fundamental observation underlying the block principal pivoting algorithm is that Equation~\ref{eqn:nnls} can be rewritten (using the definition of the $L^2$ norm) as a quadratic program:
\begin{equation}
\label{eqn:cqpr}
\min_x \mathord{-}(A^T b)^T x + \frac{1}{2} x^T A^T\!A x , \quad  \mbox{ with } x \geq 0.
\end{equation}
Since $A$ has full rank, $A^T\!A$ is positive-definite, and this is a convex program
which can be rewritten as a linear complementarity problem:
\begin{equation}
\label{eqn:lcp}
y = A^T\! A x - A^T b, \quad y \geq 0, x \geq 0, \left< x, y \right> = 0.
\end{equation}
The last condition means that the nonzero entries of $x$ and $y$ occupy complementary
variables: any given position must vanish in $x$ or $y$ (or both). In fact, the nonzero entries in $y$ represent variables in $x$ which would decrease the residual $Ax - b$ still further by becoming negative, and so are set to zero in the solution to the constrained problem. 

Suppose we have a division of the $n$ indices of the variables in $x$ into complementary sets $F$ and $G$, and let $x_F$ and $y_G$ denote pairs of vectors with the indices of their nonzero entries in these sets. Then we say that the pair $(x_F, y_G)$ is a \emph{complementary basic solution} of Equation~\ref{eqn:lcp} if $x_F$ is a solution of the unconstrained least squares problem
\begin{equation}
\label{eqn:xfprob}
\min_{x_F \in \mathbb{R}^{|F|}} \frac{1}{2} || A_F x_F - b||_2^2,
\end{equation}
where $A_F$ is formed from $A$ by selecting the columns indexed by $F$, and $y_G$ is obtained by
\begin{equation}
\label{eqn:ygeqn}
y_G = A_G^T \left( A_F x_F - b \right).
\end{equation}
If $x_F \geq 0$ and $y_G \geq 0$, then the solution is \emph{feasible}. Otherwise it is
\emph{infeasible}, and we refer to the negative entries of $x_F$ and $y_G$ as \emph{infeasible variables}. The idea of the algorithm is to proceed through infeasible complementary basic solutions of (\ref{eqn:lcp}) to the unique feasible solution by exchanging infeasible variables between $F$ and $G$ and updating $x_F$ and $y_G$ by (\ref{eqn:xfprob}) and~(\ref{eqn:ygeqn}). To minimize the number of solutions of the least-squares problem in~(\ref{eqn:xfprob}), it is desirable to exchange variables in large groups if possible. In rare cases, this may cause the algorithm to cycle. Therefore, we fall back on exchanging variables one at a time if no progress is made for a certain number of iterations with the larger exchanges.

The original block-principal pivoting algorithm works very well for what we call ``numerically nondegenerate'' problems, where each of the variables in $F$ and $G$ have values distinguishable from zero by the unconstrained solver in the feasible solution. If this is not the case, a variable with solution value close to zero may be passed back and forth between $F$ and $G$, each time reported as slightly negative due to error in the unconstrained solver. We work around this problem by zeroing variables in the unconstrained solution that are within $10^{\mathord{-}12}$ of zero. Although this strategy works well in practice, we have not developed its theoretical basis. Indeed, this seems to be an unexplored area: \cite{MR95a:90059} do not discuss the issue in their original development of the algorithm and Matstoms' \snnls implementation fails in this case.

The details are summarized below.

\begin{center}
Block principal pivoting algorithm (modified for numerically degenerate problems)\\
\bigskip

\begin{minipage}{6.5in}
\begin{tabbing}
Let $F = \emptyset$, $G = \{1, \dots, n\}$, $x = 0$, $y = -A^T b$, and $p = 3$. \\
Set $N = \infty$.\\
{\bf while} $(x_F,y_G)$ is an infeasible solution \{ \\
\hspace*{1em}Set $n$ to the number of negative entries in $x_F$ and $y_G$.\\
\hspace*{1em}{\bf if} $n < N$ (the number of infeasibles has decreased) \{ \\
\hspace*{1em}\hspace*{1em}Set $N = n$ and $p = 3$.\\
\hspace*{1em}\hspace*{1em}Exchange all infeasible variables between $F$ and $G$.\\
\hspace*{1em}\} {\bf else} \{\\
\hspace*{1em}\hspace*{1em} {\bf if} $p > 0$ \{ \\
\hspace*{1em}\hspace*{1em}\hspace*{1em}Set $p = p - 1$.\\
\hspace*{1em}\hspace*{1em}\hspace*{1em}Exchange all infeasible variables between $F$ and $G$.\\
\hspace*{1em}\hspace*{1em}\} {\bf else} \{ \\
\hspace*{1em}\hspace*{1em}\hspace*{1em}Exchange only the infeasible variable with largest index.\\
\hspace*{1em}\hspace*{1em}\} \\
\hspace*{1em} \} \\
\hspace*{1em}Update $x_F$ and $y_G$ by Equations~\ref{eqn:xfprob} and \ref{eqn:ygeqn}.\\
\hspace*{1em}Set variables in $x_F < 10^{-12}$ and $y_G < 10^{-12}$ to zero.\\
\} \\
\end{tabbing}
\end{minipage}
\end{center}

\vspace{-0.2in}
\subsection{The normal equations solver}

Solving Equation~\ref{eqn:xfprob} requires an unconstrained least-squares solver. We will often 
be able to do this by the method of normal equations. Since some of our software design choices depend on the details of this standard method, we review them here. To solve a least-squares problem $Ax = b$ using the normal equations, one solves 
\begin{equation}
A^T\!A x = A^T b
\end{equation}
using a Cholesky factorization of the symmetric matrix $A^T\!A$. This is extremely fast. For an $m \times n$ dense matrix $A$, the matrix multiplication required to form $A^T\!A$ requires $n^2 m$ flops, which is more expensive than the standard Cholesky algorithm which is known to take $\frac{1}{3} n^3 + O(n^2)$ flops. For our sparse matrix problems, we found a comparable relationship between the time required for a sparse matrix-multiply and the \texttt{TAUCS} sparse Cholesky algorithm. 

The numerical performance of this method can be a problem. The condition number of $A^T\!A$ is the square of the condition number $\kappa$ of $A$. For this reason, we must expect a relative error of about $c \kappa^2 \epsilon$, where $\epsilon$ is the machine epsilon ($\simeq \!\! 10^{-16}$ in our double-precision code), and $c$ is not large. As~\cite{MR92k:65056} points out, the Cholesky decomposition may fail entirely when $\kappa^2 \epsilon \geq 1$, so we cannot expect this method to handle matrices with $\kappa > 10^8$. Our tests indicate that this simple analysis predicts the error in the normal equations solver very well (see Section~\ref{sec:testing}), so we can anticipate the accuracy of the solver by estimating the condition number of $A^T\!A$.


\section{Software architecture}

Our primary design goal in the development of \tsnnls was to create the most efficient solver which met the user's accuracy requirements and did not depend on commercial software or restricted libraries. It is clear that the heart of the algorithm is the solution of the least-squares problem in Equation~\ref{eqn:xfprob} for the new $x_F$. But the way these solutions are used is quite interesting. In the intermediate stages of the calculation, we only use $x_F$ and $y_G$ to search for infeasible variables to shift between $F$ and $G$. So we need only calculate correct \textit{signs} for all the variables--- beyond this the numerical quality of these solutions is unimportant. But the last solution of Equations~\ref{eqn:xfprob} and~\ref{eqn:ygeqn} is the result of the algorithm, so this solution must meet the user's full accuracy needs. Our implementation takes advantage of this situation by using the method of normal equations for the intermediate solutions of Equation~\ref{eqn:xfprob} and then recomputing the final solution using the more accurate~\texttt{LSQR} solver of~\cite{355989}. 

The method of normal equations is already fast. But two of our implementation ideas improve its speed still further in our solver. As we mentioned in Section~\ref{sec:algorithms}, computing $A^{T}\!A$ is the most expensive step in the normal equations solver. A first observation is that we need not form $A^T$ explicitly in order to perform this matrix multiplication, since $A^T\!A_{ij}$ is just the dot product of the $i^{\text{th}}$ and $j^{\text{th}}$ columns of $A$. This provides some speedup. More importantly, we observe that each least-squares problem in \tsnnls is based on a submatrix $A_F$ of the same matrix~$A$. Since $A_F^T\!A_F$ is a submatrix of $A^T\!A$, we can precompute the full $A^T\!A$ and pass submatrices to the normal equations solver as required. This is a significant speed increase.
We make use of the \texttt{TAUCS} library of \cite{taucs} for highly optimized computation of the sparse Cholesky factorization needed for the method of normal equations.

We can estimate the relative error $\kappa^2 \epsilon$ of each normal equations solution by computing the condition number $\kappa^2$ of $A_F^T\!A_F$ with the \texttt{LAPACK} function \texttt{dpocon}. Since we have already computed the Cholesky factorization of $A^T\!A$ as part of the solution, this takes very little additional time in the computation. This is used to determine when a switch to a final step with \texttt{LSQR} is necessary for error control. 

In order to simplify its' use in other applications, our library incorporates simplified forms of the \TAUCS and \LSQR distributions. These are compiled directly into our library, so there is no need for the user to obtain and link with these codes separately.

\section{Software Testing}
\label{sec:testing}

We tested our implementation using problems produced by the \LSQR test generator which generates arbitrarily sized matrices with specified condition number and solution~(see \cite{355989} for details on how the generator works). We report on the relative error of our method with the problems of type $P(80,70,4,x)$, which were typical of our test results. Here $80$ and $70$ are the dimensions of the matrix, each singular value is repeated $4$ times, and $x$ is a parameter which controls the condition number of the problem. For these matrices the exact solution was known in advance, so we could measure the relative error of our solutions as a function of condition number. 

The results of this test are shown in Figure~\ref{fig:errorgraph}. The line of datapoints indicated by $\times$ shows the error in \tsnnls using only our normal equations solver. As expected, it fits very well to about $\frac{1}{10}\kappa^{2}\eps$ where $\kappa$ is the condition number of the matrix and $\epsilon$ is machine epsilon. The second set of data points (denoted by $\triangleright$) shows that we usually improve our relative error by $2$ or $3$ orders of magnitude by recomputing the final solution with $\LSQR$. The third set of data points (denoted by \parbox[b]{5pt}{\setlength{\unitlength}{0.240900pt}\begin{picture}(5,12)(0,0)\put(7,9){\circle{12}}\end{picture}}) plots the error from the \texttt{MATLAB} function \texttt{lsqnonneg} on these problems. For condition numbers up to $10^6$, we see that \tsnnls and \lsqnonneg have comparable accuracy. But surprisingly, our method seems to be more stable than \lsqnonneg for very ill-conditioned problems. 

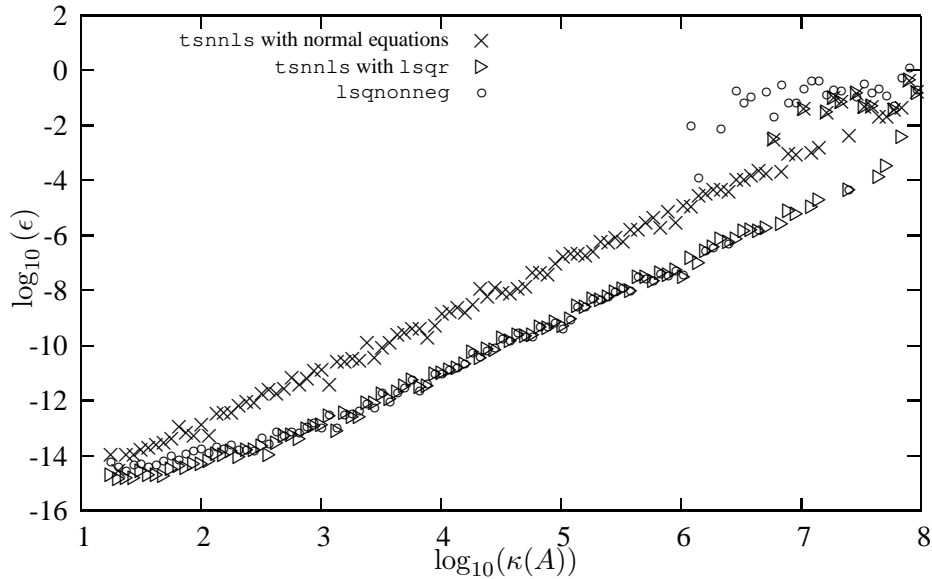
\begin{figure}
\begin{center}
\setlength{\unitlength}{0.240900pt}
\ifx\plotpoint\undefined\newsavebox{\plotpoint}\fi
\sbox{\plotpoint}{\rule[-0.200pt]{0.400pt}{0.400pt}}%
\begin{picture}(1500,900)(0,0)
\put(675,-10){$\log_{10}(\kappa(A))$}
\put(10,400){\rotatebox{90}{$\log_{10}\left( \epsilon \right)$}}

\sbox{\plotpoint}{\rule[-0.200pt]{0.400pt}{0.400pt}}%
\put(120.0,82.0){\rule[-0.200pt]{4.818pt}{0.400pt}}
\put(100,82){\makebox(0,0)[r]{-16}}
\put(1419.0,82.0){\rule[-0.200pt]{4.818pt}{0.400pt}}
\put(120.0,168.0){\rule[-0.200pt]{4.818pt}{0.400pt}}
\put(100,168){\makebox(0,0)[r]{-14}}
\put(1419.0,168.0){\rule[-0.200pt]{4.818pt}{0.400pt}}
\put(120.0,255.0){\rule[-0.200pt]{4.818pt}{0.400pt}}
\put(100,255){\makebox(0,0)[r]{-12}}
\put(1419.0,255.0){\rule[-0.200pt]{4.818pt}{0.400pt}}
\put(120.0,341.0){\rule[-0.200pt]{4.818pt}{0.400pt}}
\put(100,341){\makebox(0,0)[r]{-10}}
\put(1419.0,341.0){\rule[-0.200pt]{4.818pt}{0.400pt}}
\put(120.0,428.0){\rule[-0.200pt]{4.818pt}{0.400pt}}
\put(100,428){\makebox(0,0)[r]{-8}}
\put(1419.0,428.0){\rule[-0.200pt]{4.818pt}{0.400pt}}
\put(120.0,514.0){\rule[-0.200pt]{4.818pt}{0.400pt}}
\put(100,514){\makebox(0,0)[r]{-6}}
\put(1419.0,514.0){\rule[-0.200pt]{4.818pt}{0.400pt}}
\put(120.0,601.0){\rule[-0.200pt]{4.818pt}{0.400pt}}
\put(100,601){\makebox(0,0)[r]{-4}}
\put(1419.0,601.0){\rule[-0.200pt]{4.818pt}{0.400pt}}
\put(120.0,687.0){\rule[-0.200pt]{4.818pt}{0.400pt}}
\put(100,687){\makebox(0,0)[r]{-2}}
\put(1419.0,687.0){\rule[-0.200pt]{4.818pt}{0.400pt}}
\put(120.0,774.0){\rule[-0.200pt]{4.818pt}{0.400pt}}
\put(100,774){\makebox(0,0)[r]{ 0}}
\put(1419.0,774.0){\rule[-0.200pt]{4.818pt}{0.400pt}}
\put(120.0,860.0){\rule[-0.200pt]{4.818pt}{0.400pt}}
\put(100,860){\makebox(0,0)[r]{ 2}}
\put(1419.0,860.0){\rule[-0.200pt]{4.818pt}{0.400pt}}
\put(120.0,82.0){\rule[-0.200pt]{0.400pt}{4.818pt}}
\put(120,41){\makebox(0,0){ 1}}
\put(120.0,840.0){\rule[-0.200pt]{0.400pt}{4.818pt}}
\put(308.0,82.0){\rule[-0.200pt]{0.400pt}{4.818pt}}
\put(308,41){\makebox(0,0){ 2}}
\put(308.0,840.0){\rule[-0.200pt]{0.400pt}{4.818pt}}
\put(497.0,82.0){\rule[-0.200pt]{0.400pt}{4.818pt}}
\put(497,41){\makebox(0,0){ 3}}
\put(497.0,840.0){\rule[-0.200pt]{0.400pt}{4.818pt}}
\put(685.0,82.0){\rule[-0.200pt]{0.400pt}{4.818pt}}
\put(685,41){\makebox(0,0){ 4}}
\put(685.0,840.0){\rule[-0.200pt]{0.400pt}{4.818pt}}
\put(874.0,82.0){\rule[-0.200pt]{0.400pt}{4.818pt}}
\put(874,41){\makebox(0,0){ 5}}
\put(874.0,840.0){\rule[-0.200pt]{0.400pt}{4.818pt}}
\put(1062.0,82.0){\rule[-0.200pt]{0.400pt}{4.818pt}}
\put(1062,41){\makebox(0,0){ 6}}
\put(1062.0,840.0){\rule[-0.200pt]{0.400pt}{4.818pt}}
\put(1251.0,82.0){\rule[-0.200pt]{0.400pt}{4.818pt}}
\put(1251,41){\makebox(0,0){ 7}}
\put(1251.0,840.0){\rule[-0.200pt]{0.400pt}{4.818pt}}
\put(1439.0,82.0){\rule[-0.200pt]{0.400pt}{4.818pt}}
\put(1439,41){\makebox(0,0){ 8}}
\put(1439.0,840.0){\rule[-0.200pt]{0.400pt}{4.818pt}}
\put(120.0,82.0){\rule[-0.200pt]{317.747pt}{0.400pt}}
\put(1439.0,82.0){\rule[-0.200pt]{0.400pt}{187.420pt}}
\put(120.0,860.0){\rule[-0.200pt]{317.747pt}{0.400pt}}
\put(120.0,82.0){\rule[-0.200pt]{0.400pt}{187.420pt}}
\put(700,820){\makebox(0,0)[r]{\scriptsize{\texttt{tsnnls} with normal equations}}}
\put(168,169){\makebox(0,0){$\times$}}
\put(180,145){\makebox(0,0){$\times$}}
\put(192,169){\makebox(0,0){$\times$}}
\put(204,170){\makebox(0,0){$\times$}}
\put(215,177){\makebox(0,0){$\times$}}
\put(227,181){\makebox(0,0){$\times$}}
\put(239,186){\makebox(0,0){$\times$}}
\put(251,188){\makebox(0,0){$\times$}}
\put(263,195){\makebox(0,0){$\times$}}
\put(275,213){\makebox(0,0){$\times$}}
\put(286,204){\makebox(0,0){$\times$}}
\put(298,200){\makebox(0,0){$\times$}}
\put(310,217){\makebox(0,0){$\times$}}
\put(322,199){\makebox(0,0){$\times$}}
\put(334,234){\makebox(0,0){$\times$}}
\put(345,235){\makebox(0,0){$\times$}}
\put(357,235){\makebox(0,0){$\times$}}
\put(369,247){\makebox(0,0){$\times$}}
\put(381,253){\makebox(0,0){$\times$}}
\put(393,252){\makebox(0,0){$\times$}}
\put(405,265){\makebox(0,0){$\times$}}
\put(416,272){\makebox(0,0){$\times$}}
\put(428,266){\makebox(0,0){$\times$}}
\put(440,274){\makebox(0,0){$\times$}}
\put(452,291){\makebox(0,0){$\times$}}
\put(464,280){\makebox(0,0){$\times$}}
\put(476,289){\makebox(0,0){$\times$}}
\put(487,302){\makebox(0,0){$\times$}}
\put(499,303){\makebox(0,0){$\times$}}
\put(511,280){\makebox(0,0){$\times$}}
\put(523,316){\makebox(0,0){$\times$}}
\put(535,316){\makebox(0,0){$\times$}}
\put(547,320){\makebox(0,0){$\times$}}
\put(558,318){\makebox(0,0){$\times$}}
\put(570,345){\makebox(0,0){$\times$}}
\put(582,322){\makebox(0,0){$\times$}}
\put(594,338){\makebox(0,0){$\times$}}
\put(606,346){\makebox(0,0){$\times$}}
\put(618,358){\makebox(0,0){$\times$}}
\put(629,361){\makebox(0,0){$\times$}}
\put(641,367){\makebox(0,0){$\times$}}
\put(653,367){\makebox(0,0){$\times$}}
\put(665,353){\makebox(0,0){$\times$}}
\put(677,373){\makebox(0,0){$\times$}}
\put(688,391){\makebox(0,0){$\times$}}
\put(700,394){\makebox(0,0){$\times$}}
\put(712,401){\makebox(0,0){$\times$}}
\put(724,393){\makebox(0,0){$\times$}}
\put(736,406){\makebox(0,0){$\times$}}
\put(748,431){\makebox(0,0){$\times$}}
\put(759,418){\makebox(0,0){$\times$}}
\put(771,432){\makebox(0,0){$\times$}}
\put(783,424){\makebox(0,0){$\times$}}
\put(795,423){\makebox(0,0){$\times$}}
\put(807,430){\makebox(0,0){$\times$}}
\put(819,434){\makebox(0,0){$\times$}}
\put(830,455){\makebox(0,0){$\times$}}
\put(842,456){\makebox(0,0){$\times$}}
\put(854,452){\makebox(0,0){$\times$}}
\put(866,469){\makebox(0,0){$\times$}}
\put(878,481){\makebox(0,0){$\times$}}
\put(890,486){\makebox(0,0){$\times$}}
\put(901,486){\makebox(0,0){$\times$}}
\put(913,483){\makebox(0,0){$\times$}}
\put(925,488){\makebox(0,0){$\times$}}
\put(937,505){\makebox(0,0){$\times$}}
\put(949,503){\makebox(0,0){$\times$}}
\put(960,510){\makebox(0,0){$\times$}}
\put(972,505){\makebox(0,0){$\times$}}
\put(984,524){\makebox(0,0){$\times$}}
\put(996,524){\makebox(0,0){$\times$}}
\put(1008,534){\makebox(0,0){$\times$}}
\put(1020,542){\makebox(0,0){$\times$}}
\put(1031,526){\makebox(0,0){$\times$}}
\put(1043,551){\makebox(0,0){$\times$}}
\put(1055,535){\makebox(0,0){$\times$}}
\put(1067,561){\makebox(0,0){$\times$}}
\put(1079,560){\makebox(0,0){$\times$}}
\put(1091,577){\makebox(0,0){$\times$}}
\put(1102,580){\makebox(0,0){$\times$}}
\put(1114,586){\makebox(0,0){$\times$}}
\put(1126,586){\makebox(0,0){$\times$}}
\put(1138,583){\makebox(0,0){$\times$}}
\put(1150,602){\makebox(0,0){$\times$}}
\put(1162,602){\makebox(0,0){$\times$}}
\put(1173,608){\makebox(0,0){$\times$}}
\put(1185,616){\makebox(0,0){$\times$}}
\put(1197,612){\makebox(0,0){$\times$}}
\put(1209,665){\makebox(0,0){$\times$}}
\put(1221,615){\makebox(0,0){$\times$}}
\put(1232,643){\makebox(0,0){$\times$}}
\put(1244,641){\makebox(0,0){$\times$}}
\put(1256,713){\makebox(0,0){$\times$}}
\put(1268,644){\makebox(0,0){$\times$}}
\put(1280,652){\makebox(0,0){$\times$}}
\put(1292,707){\makebox(0,0){$\times$}}
\put(1303,730){\makebox(0,0){$\times$}}
\put(1315,724){\makebox(0,0){$\times$}}
\put(1327,671){\makebox(0,0){$\times$}}
\put(1339,737){\makebox(0,0){$\times$}}
\put(1351,716){\makebox(0,0){$\times$}}
\put(1363,717){\makebox(0,0){$\times$}}
\put(1374,701){\makebox(0,0){$\times$}}
\put(1386,701){\makebox(0,0){$\times$}}
\put(1398,711){\makebox(0,0){$\times$}}
\put(1410,715){\makebox(0,0){$\times$}}
\put(1422,758){\makebox(0,0){$\times$}}
\put(1434,740){\makebox(0,0){$\times$}}
\put(750,820){\makebox(0,0){$\times$}} 
\put(700,779){\makebox(0,0)[r]{\scriptsize{\texttt{tsnnls} with \texttt{lsqr}}}}
\put(168,138){\makebox(0,0){$\triangleright$}}
\put(180,131){\makebox(0,0){$\triangleright$}}
\put(192,133){\makebox(0,0){$\triangleright$}}
\put(204,133){\makebox(0,0){$\triangleright$}}
\put(215,143){\makebox(0,0){$\triangleright$}}
\put(227,137){\makebox(0,0){$\triangleright$}}
\put(239,138){\makebox(0,0){$\triangleright$}}
\put(251,135){\makebox(0,0){$\triangleright$}}
\put(263,146){\makebox(0,0){$\triangleright$}}
\put(275,152){\makebox(0,0){$\triangleright$}}
\put(286,148){\makebox(0,0){$\triangleright$}}
\put(298,154){\makebox(0,0){$\triangleright$}}
\put(310,154){\makebox(0,0){$\triangleright$}}
\put(322,159){\makebox(0,0){$\triangleright$}}
\put(334,171){\makebox(0,0){$\triangleright$}}
\put(345,168){\makebox(0,0){$\triangleright$}}
\put(357,175){\makebox(0,0){$\triangleright$}}
\put(369,165){\makebox(0,0){$\triangleright$}}
\put(381,177){\makebox(0,0){$\triangleright$}}
\put(393,177){\makebox(0,0){$\triangleright$}}
\put(405,183){\makebox(0,0){$\triangleright$}}
\put(416,169){\makebox(0,0){$\triangleright$}}
\put(428,187){\makebox(0,0){$\triangleright$}}
\put(440,198){\makebox(0,0){$\triangleright$}}
\put(452,200){\makebox(0,0){$\triangleright$}}
\put(464,194){\makebox(0,0){$\triangleright$}}
\put(476,210){\makebox(0,0){$\triangleright$}}
\put(487,214){\makebox(0,0){$\triangleright$}}
\put(499,216){\makebox(0,0){$\triangleright$}}
\put(511,230){\makebox(0,0){$\triangleright$}}
\put(523,206){\makebox(0,0){$\triangleright$}}
\put(535,234){\makebox(0,0){$\triangleright$}}
\put(547,229){\makebox(0,0){$\triangleright$}}
\put(558,229){\makebox(0,0){$\triangleright$}}
\put(570,250){\makebox(0,0){$\triangleright$}}
\put(582,250){\makebox(0,0){$\triangleright$}}
\put(594,264){\makebox(0,0){$\triangleright$}}
\put(606,257){\makebox(0,0){$\triangleright$}}
\put(618,268){\makebox(0,0){$\triangleright$}}
\put(629,277){\makebox(0,0){$\triangleright$}}
\put(641,286){\makebox(0,0){$\triangleright$}}
\put(653,275){\makebox(0,0){$\triangleright$}}
\put(665,277){\makebox(0,0){$\triangleright$}}
\put(677,296){\makebox(0,0){$\triangleright$}}
\put(688,297){\makebox(0,0){$\triangleright$}}
\put(700,302){\makebox(0,0){$\triangleright$}}
\put(712,305){\makebox(0,0){$\triangleright$}}
\put(724,312){\makebox(0,0){$\triangleright$}}
\put(736,330){\makebox(0,0){$\triangleright$}}
\put(748,321){\makebox(0,0){$\triangleright$}}
\put(759,335){\makebox(0,0){$\triangleright$}}
\put(771,334){\makebox(0,0){$\triangleright$}}
\put(783,353){\makebox(0,0){$\triangleright$}}
\put(795,348){\makebox(0,0){$\triangleright$}}
\put(807,358){\makebox(0,0){$\triangleright$}}
\put(819,355){\makebox(0,0){$\triangleright$}}
\put(830,358){\makebox(0,0){$\triangleright$}}
\put(842,370){\makebox(0,0){$\triangleright$}}
\put(854,369){\makebox(0,0){$\triangleright$}}
\put(866,378){\makebox(0,0){$\triangleright$}}
\put(878,372){\makebox(0,0){$\triangleright$}}
\put(890,383){\makebox(0,0){$\triangleright$}}
\put(901,403){\makebox(0,0){$\triangleright$}}
\put(913,401){\makebox(0,0){$\triangleright$}}
\put(925,414){\makebox(0,0){$\triangleright$}}
\put(937,413){\makebox(0,0){$\triangleright$}}
\put(949,417){\makebox(0,0){$\triangleright$}}
\put(960,425){\makebox(0,0){$\triangleright$}}
\put(972,430){\makebox(0,0){$\triangleright$}}
\put(984,427){\makebox(0,0){$\triangleright$}}
\put(996,449){\makebox(0,0){$\triangleright$}}
\put(1008,448){\makebox(0,0){$\triangleright$}}
\put(1020,442){\makebox(0,0){$\triangleright$}}
\put(1031,454){\makebox(0,0){$\triangleright$}}
\put(1043,452){\makebox(0,0){$\triangleright$}}
\put(1055,460){\makebox(0,0){$\triangleright$}}
\put(1067,449){\makebox(0,0){$\triangleright$}}
\put(1079,479){\makebox(0,0){$\triangleright$}}
\put(1091,470){\makebox(0,0){$\triangleright$}}
\put(1102,489){\makebox(0,0){$\triangleright$}}
\put(1114,496){\makebox(0,0){$\triangleright$}}
\put(1126,508){\makebox(0,0){$\triangleright$}}
\put(1138,503){\makebox(0,0){$\triangleright$}}
\put(1150,508){\makebox(0,0){$\triangleright$}}
\put(1162,521){\makebox(0,0){$\triangleright$}}
\put(1173,523){\makebox(0,0){$\triangleright$}}
\put(1185,521){\makebox(0,0){$\triangleright$}}
\put(1197,526){\makebox(0,0){$\triangleright$}}
\put(1209,665){\makebox(0,0){$\triangleright$}}
\put(1221,532){\makebox(0,0){$\triangleright$}}
\put(1232,552){\makebox(0,0){$\triangleright$}}
\put(1244,548){\makebox(0,0){$\triangleright$}}
\put(1256,713){\makebox(0,0){$\triangleright$}}
\put(1268,558){\makebox(0,0){$\triangleright$}}
\put(1280,569){\makebox(0,0){$\triangleright$}}
\put(1292,707){\makebox(0,0){$\triangleright$}}
\put(1303,730){\makebox(0,0){$\triangleright$}}
\put(1315,724){\makebox(0,0){$\triangleright$}}
\put(1327,585){\makebox(0,0){$\triangleright$}}
\put(1339,737){\makebox(0,0){$\triangleright$}}
\put(1351,716){\makebox(0,0){$\triangleright$}}
\put(1363,717){\makebox(0,0){$\triangleright$}}
\put(1374,605){\makebox(0,0){$\triangleright$}}
\put(1386,622){\makebox(0,0){$\triangleright$}}
\put(1398,710){\makebox(0,0){$\triangleright$}}
\put(1410,669){\makebox(0,0){$\triangleright$}}
\put(1422,758){\makebox(0,0){$\triangleright$}}
\put(1434,737){\makebox(0,0){$\triangleright$}}
\put(750,779){\makebox(0,0){$\triangleright$}} 
\sbox{\plotpoint}{\rule[-0.400pt]{0.800pt}{0.800pt}}%
\sbox{\plotpoint}{\rule[-0.200pt]{0.400pt}{0.400pt}}%
\put(700,738){\makebox(0,0)[r]{\scriptsize{\texttt{lsqnonneg}}}}
\sbox{\plotpoint}{\rule[-0.400pt]{0.800pt}{0.800pt}}%
\put(168,158){\circle{12}}
\put(180,150){\circle{12}}
\put(192,144){\circle{12}}
\put(204,154){\circle{12}}
\put(215,155){\circle{12}}
\put(227,151){\circle{12}}
\put(239,154){\circle{12}}
\put(251,159){\circle{12}}
\put(263,167){\circle{12}}
\put(275,162){\circle{12}}
\put(286,170){\circle{12}}
\put(298,175){\circle{12}}
\put(310,178){\circle{12}}
\put(322,172){\circle{12}}
\put(334,182){\circle{12}}
\put(345,179){\circle{12}}
\put(357,185){\circle{12}}
\put(369,177){\circle{12}}
\put(381,177){\circle{12}}
\put(393,176){\circle{12}}
\put(405,196){\circle{12}}
\put(416,187){\circle{12}}
\put(428,205){\circle{12}}
\put(440,199){\circle{12}}
\put(452,205){\circle{12}}
\put(464,204){\circle{12}}
\put(476,213){\circle{12}}
\put(487,219){\circle{12}}
\put(499,212){\circle{12}}
\put(511,232){\circle{12}}
\put(523,211){\circle{12}}
\put(535,232){\circle{12}}
\put(547,234){\circle{12}}
\put(558,238){\circle{12}}
\put(570,250){\circle{12}}
\put(582,243){\circle{12}}
\put(594,266){\circle{12}}
\put(606,253){\circle{12}}
\put(618,266){\circle{12}}
\put(629,275){\circle{12}}
\put(641,287){\circle{12}}
\put(653,270){\circle{12}}
\put(665,281){\circle{12}}
\put(677,296){\circle{12}}
\put(688,297){\circle{12}}
\put(700,303){\circle{12}}
\put(712,306){\circle{12}}
\put(724,312){\circle{12}}
\put(736,329){\circle{12}}
\put(748,323){\circle{12}}
\put(759,333){\circle{12}}
\put(771,338){\circle{12}}
\put(783,352){\circle{12}}
\put(795,349){\circle{12}}
\put(807,361){\circle{12}}
\put(819,355){\circle{12}}
\put(830,355){\circle{12}}
\put(842,371){\circle{12}}
\put(854,371){\circle{12}}
\put(866,376){\circle{12}}
\put(878,367){\circle{12}}
\put(890,381){\circle{12}}
\put(901,401){\circle{12}}
\put(913,402){\circle{12}}
\put(925,414){\circle{12}}
\put(937,413){\circle{12}}
\put(949,418){\circle{12}}
\put(960,426){\circle{12}}
\put(972,432){\circle{12}}
\put(984,427){\circle{12}}
\put(996,449){\circle{12}}
\put(1008,445){\circle{12}}
\put(1020,443){\circle{12}}
\put(1031,454){\circle{12}}
\put(1043,451){\circle{12}}
\put(1055,459){\circle{12}}
\put(1067,452){\circle{12}}
\put(1079,686){\circle{12}}
\put(1091,604){\circle{12}}
\put(1102,490){\circle{12}}
\put(1114,495){\circle{12}}
\put(1126,681){\circle{12}}
\put(1138,501){\circle{12}}
\put(1150,741){\circle{12}}
\put(1162,723){\circle{12}}
\put(1173,732){\circle{12}}
\put(1185,522){\circle{12}}
\put(1197,739){\circle{12}}
\put(1209,701){\circle{12}}
\put(1221,751){\circle{12}}
\put(1232,723){\circle{12}}
\put(1244,723){\circle{12}}
\put(1256,745){\circle{12}}
\put(1268,757){\circle{12}}
\put(1280,757){\circle{12}}
\put(1292,735){\circle{12}}
\put(1303,743){\circle{12}}
\put(1315,741){\circle{12}}
\put(1327,586){\circle{12}}
\put(1339,731){\circle{12}}
\put(1351,752){\circle{12}}
\put(1363,738){\circle{12}}
\put(1374,745){\circle{12}}
\put(1386,734){\circle{12}}
\put(1398,717){\circle{12}}
\put(1410,761){\circle{12}}
\put(1422,778){\circle{12}}
\put(1434,743){\circle{12}}
\put(750,738){\circle{12}} 
\sbox{\plotpoint}{\rule[-0.200pt]{0.400pt}{0.400pt}}%
\put(120.0,82.0){\rule[-0.200pt]{317.747pt}{0.400pt}}
\put(1439.0,82.0){\rule[-0.200pt]{0.400pt}{187.420pt}}
\put(120.0,860.0){\rule[-0.200pt]{317.747pt}{0.400pt}}
\put(120.0,82.0){\rule[-0.200pt]{0.400pt}{187.420pt}}
\end{picture}

\end{center}
\caption[Error in \texttt{tsnnls} and \texttt{lsqnonneg}]{This plot shows the relative error in the solution of a selection of $80 \times 70$ test problems generated by the \texttt{LSQR} test generator with inputs $P(80,70,4,x)$ and condition numbers varying from $10^1$ to $10^8$. The logarithm (base $10$) of this error $\epsilon$ is plotted against the logarithm of condition number for three codes: \tsnnls restricted to use only the normal equations solver, the final version of \texttt{tsnnls}, which recomputes the final solution with \texttt{LSQR}, and the \texttt{MATLAB} function \texttt{lsqnonneg}.}
\label{fig:errorgraph}
\end{figure} 

We also tested the performance of our software against that of \lsqnonneg and that of
the \snnls code of~\cite{sls}. All of our timing tests were performed on a dual 2.0 GHz Power Macintosh G5 running Mac OS X 10.3, compiling with \texttt{gcc\,3.3} and \texttt{-O3}, 
and linking with Apple's optimized versions of \texttt{LAPACK} and \texttt{BLAS}. We ran \snnls under~\texttt{MATLAB\,7} with argument \texttt{-nojvm}. 

We were required to make two modifications to the \snnls code to complete our tests. 
First, the \texttt{snnls} code uses the column minimum degree permutation (\texttt{colmmd}) before performing sparse Cholesky decompositions. However, as this ordering is deprecated in \texttt{MATLAB\,7} in favor of absolute minimum degree ordering, we tested against a modified \texttt{snnls} using \texttt{colamd}. This was a strict performance improvement for our test cases. We also made the same workaround to handle degenerate problems that we discussed for \tsnnls in Section~\ref{sec:algorithms}.

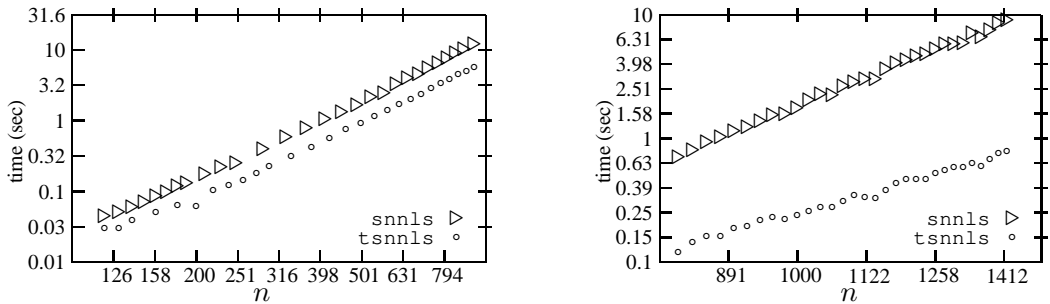
\begin{figure}

\begin{center}
\begin{minipage}{3.2in}
\setlength{\unitlength}{0.12045pt}
\ifx\plotpoint\undefined\newsavebox{\plotpoint}\fi
\begin{picture}(1500,900)(0,0)
\sbox{\plotpoint}{\rule[-0.200pt]{0.400pt}{0.400pt}}%
\put(120,82){\makebox(0,0)[r]{\scriptsize{0.01}}}
\put(140.0,193.0){\rule[-0.200pt]{4.818pt}{0.400pt}}
\put(120,193){\makebox(0,0)[r]{\scriptsize{0.03}}}
\put(1419.0,193.0){\rule[-0.200pt]{4.818pt}{0.400pt}}
\put(140.0,304.0){\rule[-0.200pt]{4.818pt}{0.400pt}}
\put(120,304){\makebox(0,0)[r]{\scriptsize{0.1}}}
\put(1419.0,304.0){\rule[-0.200pt]{4.818pt}{0.400pt}}
\put(140.0,415.0){\rule[-0.200pt]{4.818pt}{0.400pt}}
\put(120,415){\makebox(0,0)[r]{\scriptsize{0.32}}}
\put(1419.0,415.0){\rule[-0.200pt]{4.818pt}{0.400pt}}
\put(140.0,527.0){\rule[-0.200pt]{4.818pt}{0.400pt}}
\put(120,527){\makebox(0,0)[r]{ \scriptsize{1}}}
\put(1419.0,527.0){\rule[-0.200pt]{4.818pt}{0.400pt}}
\put(140.0,638.0){\rule[-0.200pt]{4.818pt}{0.400pt}}
\put(120,638){\makebox(0,0)[r]{ \scriptsize{3.2}}}
\put(1419.0,638.0){\rule[-0.200pt]{4.818pt}{0.400pt}}
\put(140.0,749.0){\rule[-0.200pt]{4.818pt}{0.400pt}}
\put(120,749){\makebox(0,0)[r]{ \scriptsize{10}}}
\put(1419.0,749.0){\rule[-0.200pt]{4.818pt}{0.400pt}}
\put(120,860){\makebox(0,0)[r]{ \scriptsize{31.6}}}
\put(140.0,82.0){\rule[-0.200pt]{0.400pt}{4.818pt}}
\put(270.0,82.0){\rule[-0.200pt]{0.400pt}{4.818pt}}
\put(270,41){\makebox(0,0){ \scriptsize{126}}}
\put(270.0,840.0){\rule[-0.200pt]{0.400pt}{4.818pt}}
\put(400.0,82.0){\rule[-0.200pt]{0.400pt}{4.818pt}}
\put(400,41){\makebox(0,0){ \scriptsize{158} }}
\put(400.0,840.0){\rule[-0.200pt]{0.400pt}{4.818pt}}
\put(530.0,82.0){\rule[-0.200pt]{0.400pt}{4.818pt}}
\put(530,41){\makebox(0,0){ \scriptsize{200}}}
\put(530.0,840.0){\rule[-0.200pt]{0.400pt}{4.818pt}}
\put(660.0,82.0){\rule[-0.200pt]{0.400pt}{4.818pt}}
\put(660,41){\makebox(0,0){ \scriptsize{251}}}
\put(660.0,840.0){\rule[-0.200pt]{0.400pt}{4.818pt}}
\put(790.0,82.0){\rule[-0.200pt]{0.400pt}{4.818pt}}
\put(790,41){\makebox(0,0){ \scriptsize{316}}}
\put(790.0,840.0){\rule[-0.200pt]{0.400pt}{4.818pt}}
\put(919.0,82.0){\rule[-0.200pt]{0.400pt}{4.818pt}}
\put(919,41){\makebox(0,0){ \scriptsize{398}}}
\put(919.0,840.0){\rule[-0.200pt]{0.400pt}{4.818pt}}
\put(1049.0,82.0){\rule[-0.200pt]{0.400pt}{4.818pt}}
\put(1049,41){\makebox(0,0){ \scriptsize{501}}}
\put(1049.0,840.0){\rule[-0.200pt]{0.400pt}{4.818pt}}
\put(1179.0,82.0){\rule[-0.200pt]{0.400pt}{4.818pt}}
\put(1179,41){\makebox(0,0){ \scriptsize{631} }}
\put(1179.0,840.0){\rule[-0.200pt]{0.400pt}{4.818pt}}
\put(1309.0,82.0){\rule[-0.200pt]{0.400pt}{4.818pt}}
\put(1309,41){\makebox(0,0){ \scriptsize{794}}}

\put(140.0,82.0){\rule[-0.200pt]{156.4645pt}{0.400pt}}
\put(1439.0,82.0){\rule[-0.200pt]{0.400pt}{93.71pt}}
\put(140.0,860.0){\rule[-0.200pt]{156.4645pt}{0.400pt}}
\put(140.0,82.0){\rule[-0.200pt]{0.400pt}{93.71pt}}

\put(1279,160){\makebox(0,0)[r]{\scriptsize{\texttt{tsnnls}}}}
\put(1349,160){\raisebox{0.0pt}{\circle{20}}}

\put(243,191){\raisebox{0.0pt}{\circle{20}}}
\put(288,191){\raisebox{0.0pt}{\circle{20}}}
\put(330,216){\raisebox{0.0pt}{\circle{20}}}
\put(405,239){\raisebox{0.0pt}{\circle{20}}}
\put(472,261){\raisebox{0.0pt}{\circle{20}}}
\put(531,260){\raisebox{0.0pt}{\circle{20}}}
\put(585,308){\raisebox{0.0pt}{\circle{20}}}
\put(634,324){\raisebox{0.0pt}{\circle{20}}}
\put(679,339){\raisebox{0.0pt}{\circle{20}}}
\put(721,361){\raisebox{0.0pt}{\circle{20}}}
\put(760,383){\raisebox{0.0pt}{\circle{20}}}
\put(830,416){\raisebox{0.0pt}{\circle{20}}}
\put(893,445){\raisebox{0.0pt}{\circle{20}}}
\put(950,473){\raisebox{0.0pt}{\circle{20}}}
\put(1001,499){\raisebox{0.0pt}{\circle{20}}}
\put(1048,520){\raisebox{0.0pt}{\circle{20}}}
\put(1091,540){\raisebox{0.0pt}{\circle{20}}}
\put(1132,561){\raisebox{0.0pt}{\circle{20}}}
\put(1169,580){\raisebox{0.0pt}{\circle{20}}}
\put(1205,596){\raisebox{0.0pt}{\circle{20}}}
\put(1238,612){\raisebox{0.0pt}{\circle{20}}}
\put(1269,630){\raisebox{0.0pt}{\circle{20}}}
\put(1299,646){\raisebox{0.0pt}{\circle{20}}}
\put(1327,657){\raisebox{0.0pt}{\circle{20}}}
\put(1354,673){\raisebox{0.0pt}{\circle{20}}}
\put(1380,682){\raisebox{0.0pt}{\circle{20}}}
\put(1404,695){\raisebox{0.0pt}{\circle{20}}}

\put(1279,220){\makebox(0,0)[r]{\scriptsize{\texttt{snnls}}}}
\put(1349,220){\makebox(0,0){$\triangleright$}}

\put(243,225){\makebox(0,0){$\triangleright$}}
\put(288,239){\makebox(0,0){$\triangleright$}}
\put(330,253){\makebox(0,0){$\triangleright$}}
\put(369,271){\makebox(0,0){$\triangleright$}}
\put(405,289){\makebox(0,0){$\triangleright$}}
\put(439,301){\makebox(0,0){$\triangleright$}}
\put(472,321){\makebox(0,0){$\triangleright$}}
\put(502,330){\makebox(0,0){$\triangleright$}}
\put(559,357){\makebox(0,0){$\triangleright$}}
\put(610,380){\makebox(0,0){$\triangleright$}}
\put(657,394){\makebox(0,0){$\triangleright$}}
\put(741,436){\makebox(0,0){$\triangleright$}}
\put(814,473){\makebox(0,0){$\triangleright$}}
\put(878,502){\makebox(0,0){$\triangleright$}}
\put(936,530){\makebox(0,0){$\triangleright$}}
\put(989,553){\makebox(0,0){$\triangleright$}}
\put(1037,574){\makebox(0,0){$\triangleright$}}
\put(1081,599){\makebox(0,0){$\triangleright$}}
\put(1122,613){\makebox(0,0){$\triangleright$}}
\put(1160,640){\makebox(0,0){$\triangleright$}}
\put(1196,661){\makebox(0,0){$\triangleright$}}
\put(1230,673){\makebox(0,0){$\triangleright$}}
\put(1261,690){\makebox(0,0){$\triangleright$}}
\put(1292,708){\makebox(0,0){$\triangleright$}}
\put(1320,724){\makebox(0,0){$\triangleright$}}
\put(1347,739){\makebox(0,0){$\triangleright$}}
\put(1373,751){\makebox(0,0){$\triangleright$}}
\put(1404,768){\makebox(0,0){$\triangleright$}}

\put(710,-40){$n$}
\put(-60,330){\rotatebox{90}{\scriptsize{time (sec)}}}

\end{picture}
\end{minipage}
\hspace{-0.2in}
\begin{minipage}{3.2in}
\setlength{\unitlength}{0.12045pt} 
\ifx\plotpoint\undefined\newsavebox{\plotpoint}\fi
\begin{picture}(1500,900)(0,0)
\sbox{\plotpoint}{\rule[-0.200pt]{0.400pt}{0.400pt}}%
\put(120,82){\makebox(0,0)[r]{\scriptsize{0.1}}}
\put(140.0,160.0){\rule[-0.200pt]{4.818pt}{0.400pt}}
\put(120,160){\makebox(0,0)[r]{\scriptsize{0.15}}}
\put(1319.0,160.0){\rule[-0.200pt]{4.818pt}{0.400pt}}
\put(140.0,238.0){\rule[-0.200pt]{4.818pt}{0.400pt}}
\put(120,238){\makebox(0,0)[r]{\scriptsize{0.25}}}
\put(1319.0,238.0){\rule[-0.200pt]{4.818pt}{0.400pt}}
\put(140.0,315.0){\rule[-0.200pt]{4.818pt}{0.400pt}}
\put(120,315){\makebox(0,0)[r]{\scriptsize{0.39}}}
\put(1319.0,315.0){\rule[-0.200pt]{4.818pt}{0.400pt}}
\put(140.0,393.0){\rule[-0.200pt]{4.818pt}{0.400pt}}
\put(120,393){\makebox(0,0)[r]{\scriptsize{0.63}}}
\put(1319.0,393.0){\rule[-0.200pt]{4.818pt}{0.400pt}}
\put(140.0,471.0){\rule[-0.200pt]{4.818pt}{0.400pt}}
\put(120,471){\makebox(0,0)[r]{ \scriptsize{1}}}
\put(1319.0,471.0){\rule[-0.200pt]{4.818pt}{0.400pt}}
\put(140.0,549.0){\rule[-0.200pt]{4.818pt}{0.400pt}}
\put(120,549){\makebox(0,0)[r]{ \scriptsize{1.58}}}
\put(1319.0,549.0){\rule[-0.200pt]{4.818pt}{0.400pt}}
\put(140.0,627.0){\rule[-0.200pt]{4.818pt}{0.400pt}}
\put(120,627){\makebox(0,0)[r]{ \scriptsize{2.51}}}
\put(1319.0,627.0){\rule[-0.200pt]{4.818pt}{0.400pt}}
\put(140.0,704.0){\rule[-0.200pt]{4.818pt}{0.400pt}}
\put(120,704){\makebox(0,0)[r]{ \scriptsize{3.98}}}
\put(1319.0,704.0){\rule[-0.200pt]{4.818pt}{0.400pt}}
\put(140.0,782.0){\rule[-0.200pt]{4.818pt}{0.400pt}}
\put(120,782){\makebox(0,0)[r]{ \scriptsize{6.31}}}
\put(1319.0,782.0){\rule[-0.200pt]{4.818pt}{0.400pt}}
\put(120,860){\makebox(0,0)[r]{ \scriptsize{10}}}
\put(356.0,82.0){\rule[-0.200pt]{0.400pt}{4.818pt}}
\put(356,41){\makebox(0,0){ \scriptsize{891}}}
\put(356.0,840.0){\rule[-0.200pt]{0.400pt}{4.818pt}}
\put(573.0,82.0){\rule[-0.200pt]{0.400pt}{4.818pt}}
\put(573,41){\makebox(0,0){ \scriptsize{1000}}}
\put(573.0,840.0){\rule[-0.200pt]{0.400pt}{4.818pt}}
\put(789.0,82.0){\rule[-0.200pt]{0.400pt}{4.818pt}}
\put(789,41){\makebox(0,0){ \scriptsize{1122}}}
\put(789.0,840.0){\rule[-0.200pt]{0.400pt}{4.818pt}}
\put(1006.0,82.0){\rule[-0.200pt]{0.400pt}{4.818pt}}
\put(1006,41){\makebox(0,0){ \scriptsize{1258}}}
\put(1006.0,840.0){\rule[-0.200pt]{0.400pt}{4.818pt}}
\put(1222.0,82.0){\rule[-0.200pt]{0.400pt}{4.818pt}}
\put(1222,41){\makebox(0,0){ \scriptsize{1412}}}
\put(1222.0,840.0){\rule[-0.200pt]{0.400pt}{4.818pt}}

\put(140.0,82.0){\rule[-0.200pt]{144.4195pt}{0.400pt}}
\put(1339.0,82.0){\rule[-0.200pt]{0.400pt}{93.71pt}} 
\put(140.0,860.0){\rule[-0.200pt]{144.4195pt}{0.400pt}} 
\put(140.0,82.0){\rule[-0.200pt]{0.400pt}{93.71pt}}

\put(1179,160){\makebox(0,0)[r]{\scriptsize{\texttt{tsnnls}}}}
\put(1249,160){\raisebox{0.0pt}{\circle{20}}}
\put(710,-40){$n$}
\put(-60,330){\rotatebox{90}{\scriptsize{time (sec)}}}

\put(200,113){\raisebox{0.0pt}{\circle{20}}}
\put(245,145){\raisebox{0.0pt}{\circle{20}}}
\put(289,165){\raisebox{0.0pt}{\circle{20}}}
\put(333,164){\raisebox{0.0pt}{\circle{20}}}
\put(375,188){\raisebox{0.0pt}{\circle{20}}}
\put(416,195){\raisebox{0.0pt}{\circle{20}}}
\put(457,215){\raisebox{0.0pt}{\circle{20}}}
\put(496,224){\raisebox{0.0pt}{\circle{20}}}
\put(535,218){\raisebox{0.0pt}{\circle{20}}}
\put(573,231){\raisebox{0.0pt}{\circle{20}}}
\put(610,242){\raisebox{0.0pt}{\circle{20}}}
\put(647,254){\raisebox{0.0pt}{\circle{20}}}
\put(683,257){\raisebox{0.0pt}{\circle{20}}}
\put(718,273){\raisebox{0.0pt}{\circle{20}}}
\put(752,292){\raisebox{0.0pt}{\circle{20}}}
\put(786,287){\raisebox{0.0pt}{\circle{20}}}
\put(819,285){\raisebox{0.0pt}{\circle{20}}}
\put(852,308){\raisebox{0.0pt}{\circle{20}}}
\put(884,330){\raisebox{0.0pt}{\circle{20}}}
\put(916,343){\raisebox{0.0pt}{\circle{20}}}
\put(947,345){\raisebox{0.0pt}{\circle{20}}}
\put(978,345){\raisebox{0.0pt}{\circle{20}}}
\put(1008,363){\raisebox{0.0pt}{\circle{20}}}
\put(1037,372){\raisebox{0.0pt}{\circle{20}}}
\put(1066,381){\raisebox{0.0pt}{\circle{20}}}
\put(1095,380){\raisebox{0.0pt}{\circle{20}}}
\put(1123,394){\raisebox{0.0pt}{\circle{20}}}
\put(1151,385){\raisebox{0.0pt}{\circle{20}}}
\put(1179,405){\raisebox{0.0pt}{\circle{20}}}
\put(1206,424){\raisebox{0.0pt}{\circle{20}}}
\put(1232,431){\raisebox{0.0pt}{\circle{20}}}

\put(1179,220){\makebox(0,0)[r]{\scriptsize{\texttt{snnls}}}}
\put(1249,220){\makebox(0,0){$\triangleright$}}

\put(200,413){\makebox(0,0){$\triangleright$}}
\put(245,433){\makebox(0,0){$\triangleright$}}
\put(289,459){\makebox(0,0){$\triangleright$}}
\put(333,474){\makebox(0,0){$\triangleright$}}
\put(375,494){\makebox(0,0){$\triangleright$}}
\put(416,507){\makebox(0,0){$\triangleright$}}
\put(457,526){\makebox(0,0){$\triangleright$}}
\put(496,543){\makebox(0,0){$\triangleright$}}
\put(535,546){\makebox(0,0){$\triangleright$}}
\put(573,564){\makebox(0,0){$\triangleright$}}
\put(610,590){\makebox(0,0){$\triangleright$}}
\put(647,611){\makebox(0,0){$\triangleright$}}
\put(683,606){\makebox(0,0){$\triangleright$}}
\put(718,635){\makebox(0,0){$\triangleright$}}
\put(752,647){\makebox(0,0){$\triangleright$}}
\put(786,656){\makebox(0,0){$\triangleright$}}
\put(819,658){\makebox(0,0){$\triangleright$}}
\put(852,689){\makebox(0,0){$\triangleright$}}
\put(884,707){\makebox(0,0){$\triangleright$}}
\put(916,717){\makebox(0,0){$\triangleright$}}
\put(947,728){\makebox(0,0){$\triangleright$}}
\put(978,734){\makebox(0,0){$\triangleright$}}
\put(1008,752){\makebox(0,0){$\triangleright$}}
\put(1037,767){\makebox(0,0){$\triangleright$}}
\put(1066,768){\makebox(0,0){$\triangleright$}}
\put(1095,771){\makebox(0,0){$\triangleright$}}
\put(1123,801){\makebox(0,0){$\triangleright$}}
\put(1151,788){\makebox(0,0){$\triangleright$}}
\put(1179,810){\makebox(0,0){$\triangleright$}}
\put(1206,834){\makebox(0,0){$\triangleright$}}
\put(1232,843){\makebox(0,0){$\triangleright$}}
\end{picture}
\end{minipage}
\end{center}
\caption[Runtime of \texttt{tsnnls} and \texttt{snnls}]{These log-log scaled plots show the runtime of \tsnnls and \snnls on density $1.0$ (left) and $0.01$ (right) matrices of size $n \times (n-10)$ on a 2.0 Ghz Apple PowerMac G5. We can see that the runtime of \tsnnls is basically proportional to that of \texttt{snnls}, but that the constant of proportionality depends on the density of the test matrices. This effect is explained below. All runtimes were calculated by repeating the test problems until the total time measured was several seconds or more.}
\label{fig:performance}
\end{figure}

Our performance results are shown in Figure~\ref{fig:performance}. We tested runtimes for randomly generated, well-conditioned matrices from \texttt{MATLAB}'s \texttt{sprandn} function. The matrices were of size $n \times (n-10)$. The plot shows runtime results for a set of density $1$ matrices and a set of density $0.01$ matrices, intended to represent general dense and sparse matrices. Each data point represents the average runtime for $10$ different matrices of the same size and density.

We can see that the runtime of our implementation is approximately proportional to that of \texttt{snnls}, and that for dense problems it is several times faster. We were surprised to note that the constant of proportionality decreases for sparse matrices and that our method is almost $10$ times faster than \snnls for matrices of density $0.01$. 

The runtime of each code is controlled by three computations: the matrix-multiply used to form $A^T\!A$, the Cholesky decomposition of that matrix, and the final recalculation of the solution (if performed). We expected to be several times faster than \snnls since our caching strategy for $A^T\!A$ eliminates a matrix-multiply operation for each pivot. The number of pivots, however, does not seem to vary with the density of our random test matrices and so does not explain our additional speed increase for sparse problems.

We explored this phenomenon by profiling both our code and \texttt{snnls}. For our random test problems at density $0.01$, the final unconstrained solution in \snnls (computed using the the \texttt{MATLAB} \verb=\= operation) consumes almost $50\%$ of the total runtime. On the other hand, in \tsnnls the final unconstrained solution (using \texttt{LSQR}) consumes only $5\%$ of runtime. Since the Cholesky decompositions take comparable time, this would seem to explain the runtime disparity.

We did not show performance data for \texttt{MATLAB}'s built-in \texttt{lsqnonneg} because it was so much slower than both \tsnnls and \texttt{snnls}. For sparse matrices, this is in part because \texttt{lsqnonneg} is a dense-matrix code. Yet, even on dense matrices, both methods outperformed \texttt{lsqnonneg} by an overwhelming amount. For instance, for a $500 \times 490$ dense matrix, \texttt{lsqrnonneg} takes over $100$ seconds to complete while \snnls and \tsnnls both finish in less than one second. We take this as a confirmation of the suggestion in MathWorks' documentation of \texttt{lsqnonneg} that it is not appropriate for large problems.

\section{Acknowledgements}

This work was funded by the National Science Foundation through the University of Georgia VIGRE grant DMS-00-8992 and DMS-02-04826 (to Cantarella and Fu). Piatek acknowledges support through DMS-0311010 to Eric Rawdon. We would like to thank our colleagues for many helpful discussions.

\bibliography{taucs_lsqr_paper,drl}

\end{document}